\documentclass[epj]{svjour}
\usepackage{graphicx}
\usepackage{color}

\begin{document}
\title{Unbiased interpolated neutron-star EoS at finite $T$ \\
for modified gravity studies}
\author{Eva Lope-Oter\inst{1} \and Felipe J. Llanes-Estrada\inst{1}
}                     
\institute{Depto. Fisica Teorica and IPARCOS, Univ. Complutense de Madrid, Plaza de las Ciencias 1, 28040 Madrid, Spain}
\date{Received: July 31st 2021 / Revised version: date}
%
\abstract{
Neutron stars and their mergers provide the highest-density regime in which Einstein's equations in full (with a matter source) can be tested against modified theories of gravity. But doing so requires {\it a priori} knowledge of the Equation of State from nuclear and hadron physics, where no contamination from computations of astrophysics observables within General Relativity has been built in. We extend the nEoS uncertainty bands, useful for this very purpose, to finite (but small) temperatures up to 30 MeV, given that the necessary computations in ChPT and in pQCD at already available in the literature.
The temperature-dependent band boundaries will be provided through the COMPOSE repository and our own website.
} 
\authorrunning{Lope-Oter and Llanes-Estrada}
\titlerunning{Unbiased $N$-star thermal EoS for modified gravity }
\maketitle
\section{Eqn. of State of Neutron Matter at finite $T$} \label{intro}

The Equation of State of neutron star matter is the basic microscopic input necessary for multiple computations of interest in astrophysics~\cite{Llanes-Estrada:2019wmz}.
At low, nuclear densities it is well constrained from laboratory data~\cite{Greif:2020pju}, but there is no universally accepted theory pathway for intermediate densities at the 100 MeV scale and above.

Numerous models with various matter contents have been deployed for this intermediate density region, for example including hyperons~\cite{Stone:2021ngh}, hybrid approaches~\cite{Benic:2014jia} including
quark matter~\cite{Ferreira:2020kvu}, employing the holographic conjecture linking QCD to a gravity dual~\cite{Jokela:2018ers}, and more.

A common approach is to employ astrophysics observables to constrain the equation of state, that therefore contains information from nuclear and particle physics, but also of General Relativity (GR), used to eliminate parts of the otherwise allowed parameter space~\cite{Tolos:2016hhl,Chatziioannou:2020pqz,Burgio:2021vgk,Tang:2021snt}. 
The inconvenience of this method is that the general relativistic interpretation of the astrophysical observables employed to constrain the equation of state becomes entrained in the EoS; thereafter, it is not safe to employ such equation to constrain modifications of GR. 

For example, there is great interest in constraining $f(R)$ theories and variations thereof~\cite{Resco:2016upv,Nashed:2021sji,Lobato:2020fxt}, or closely related scalarizations~\cite{Kruger:2021yay,Doneva:2017duq,Blazquez-Salcedo:2020ibb}, or whether the parameters of GR are sensitive to a large stress-energy-momentum density~\cite{Dobado:2011gd}, or pursue nonlocal gravity with compact objects~\cite{Panotopoulos:2021rih} among many possibilities.

But the band of allowed EoS compatible with these theories is presumably larger (since they have additional parameter freedom in the gravity sector) than in General Relativity. 
Thus, employing reduced uncertainty EoS-bands incorporating astrophysical observables leads to improper constrains on the beyond-GR theories.

A separate question, that we do not address, is how to avoid assuming General Relativity at short scales in the interpretation of the low-energy nuclear data themselves (for example, Fisher and Carlson~\cite{Carlson} have looked at constraining Non-Minimally Coupled Gravity from nuclear properties). In this work we adopt the line of thought that microscopic nuclear physics can be read-off an inertial frame with negligible tidal forces, and thus only the large accumulation of mass at neutron-star scales can cause separations from General Relativity (in a large region of intense stress-energy $T_{\mu\nu}$ and intense curvature).

For such application it seems a necessity to provide EoS that are state of the art from the point of view of nuclear and particle physics, but that are free of astrophysical bias, if they are to be used to employ neutron star data for constraining beyond-GR theories. We have provided precisely such family of Equations of State, with controlled theoretical uncertainty, relying only on first principle approaches (causality, thermodynamic stability) and perturbation theory in the appropriate density domains (chiral perturbation theory, perturbative QCD) in the nEoS sets~\cite{Oter:2019kig} (that can be downloaded from \\
{\tt https://teorica.fis.ucm.es/nEoS/}).
Meanwhile, the detection of a neutron-star merger and the spur of improved fits to ejecta~\cite{Bauswein:2019juq} and simulations of the matter-remains after the collision that followed
~\cite{DePietri:2019mti,Raithel:2021hye}
have made the need for modern EoS at finite temperature more poignant. 
We thus look at extending those sets to finite-$T$.

\section{Low density ChPT input} \label{ChPT}

Whereas there are numerous equations of state for neutron stars  extracted from chiral perturbation theory at zero temperature, full-temperature calculations do not abound. 

\begin{figure}[b]
\resizebox{0.45\textwidth}{!}{
\includegraphics{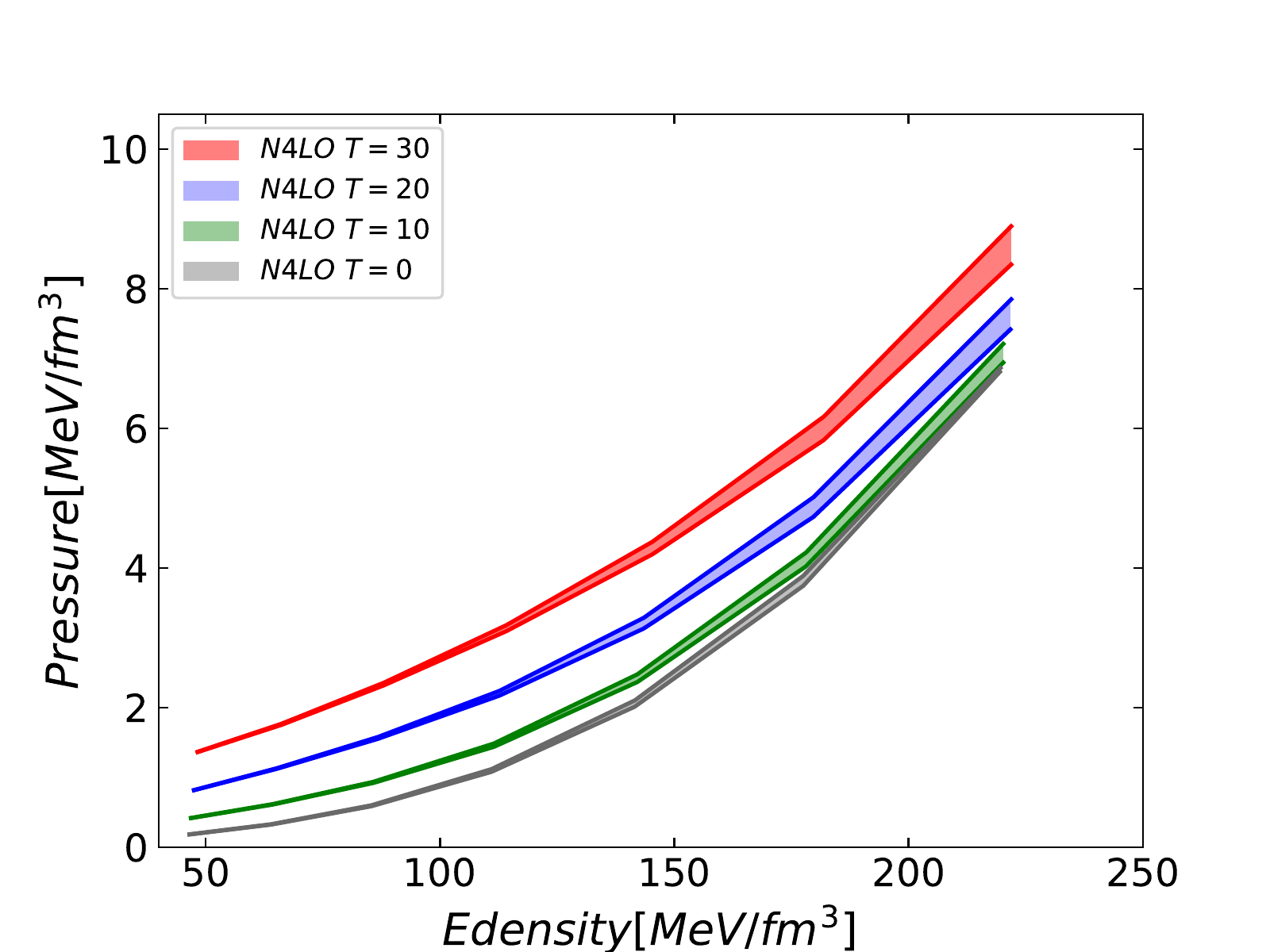}
}
\caption{\label{fig:chpt}
Equation of State of pure neutron matter for low density, taken from the ChPT computations of the nuclear potential~\cite{Sammarruca:2020urk,Sammarruca:2021mhv,Sammarruca:2016ajl}, with $T$ increasing from bottom to top in 10 MeV steps. As the number density rises, the uncertainty grows and the bands do overlap, so that the effect of temperatures much smaller than the densities is less visible.
}
\end{figure}

In this work we have leaned  on the computation of the Idaho group~\cite{Sammarruca:2020urk} for the low-density, chiral-Lagrangian tail of the computation. Their obtained Helmholtz potential depends on a $N^4LO$ treatment of the nucleon-nucleon potential; the three-body $NNN$ potential is only partially treated. Their computation is for pure neutron matter alone, and this is not a bad approximation for the low to moderate densities where the approach is valid.
The authors provide the Helmholtz free energy per nucleon,
\begin{equation}\label{UtoF}
\frac{F}{A} = \frac{E}{A} -T\frac{S}{A}
\end{equation}
from which $P$ follows directly ({\it e.g.} \cite{Koliogiannis:2021qme} for discussion)
\begin{equation}
P=- \left. \frac{\partial F}{\partial V}\right|_{N,T} \ .
\end{equation}
The convergence of the order by order chiral computations is very nice, and the effect of the temperature, interestingly, is seen to be opposite in $E/A$ and $F/A$.

In fact, Sammarruca, Machleidt and Millerson~\cite{Sammarruca:2020urk} find that $E/A$ increases with the temperature at fixed density, probably reflecting the fact that $T$ softens the step of the Fermi-Dirac occupation function, upgrading some neutrons from below to above the Fermi sea level of cold matter. Since neutrons of higher momentum have larger kinetic energy and see stronger (repulsive) chiral interactions, it is natural to expect higher energy.
However, this is compensated by the second term of Eq.~(\ref{UtoF}), so that the free energy is actually decreasing with temperature. Its derivative, however, which is the relevant quantity to compute the pressure, is not as sensitive to small $T$ except at very low densities.

\begin{figure}[h]
\resizebox{0.45\textwidth}{!}{
\includegraphics{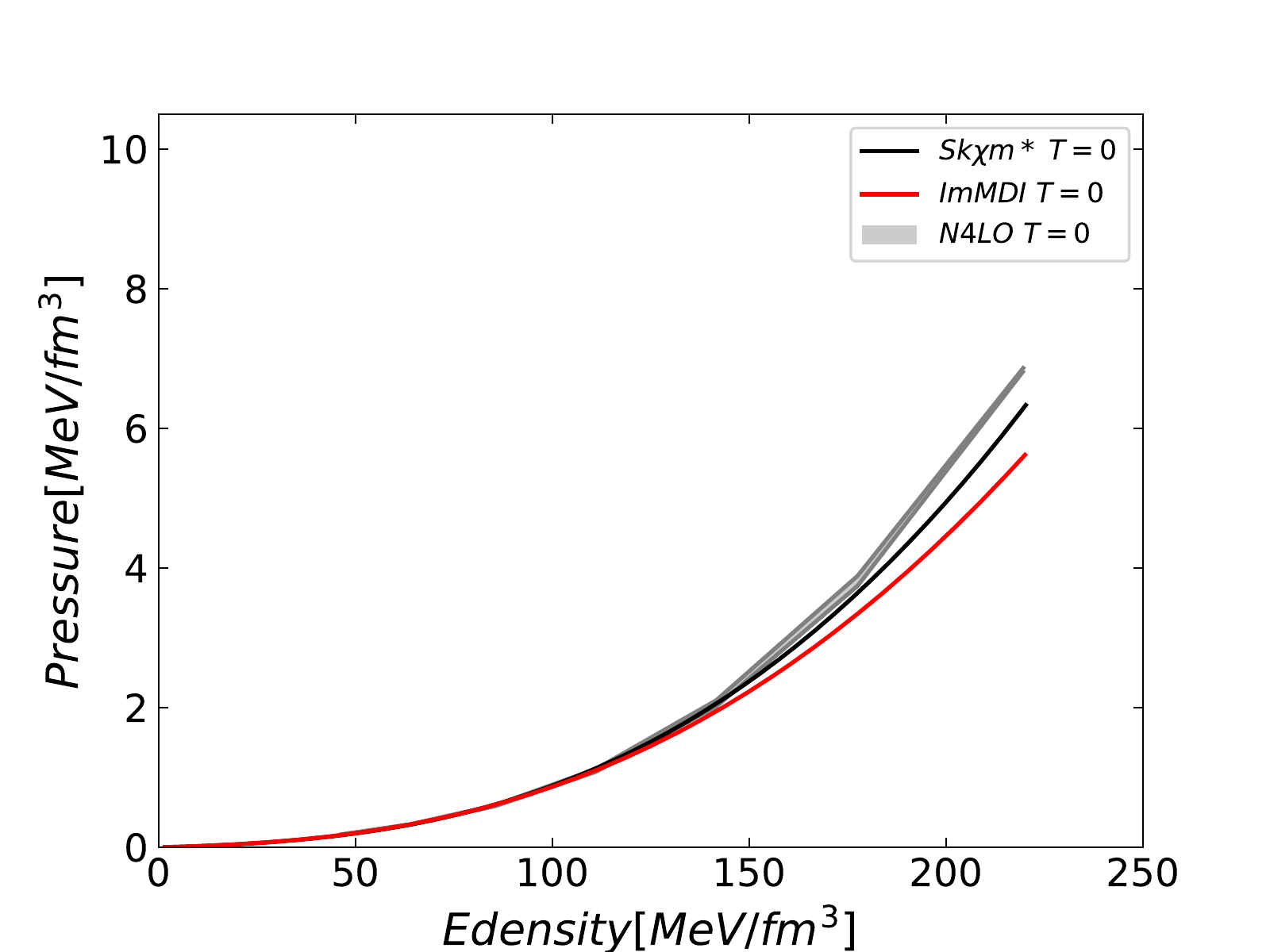}}
\resizebox{0.45\textwidth}{!}{
\includegraphics{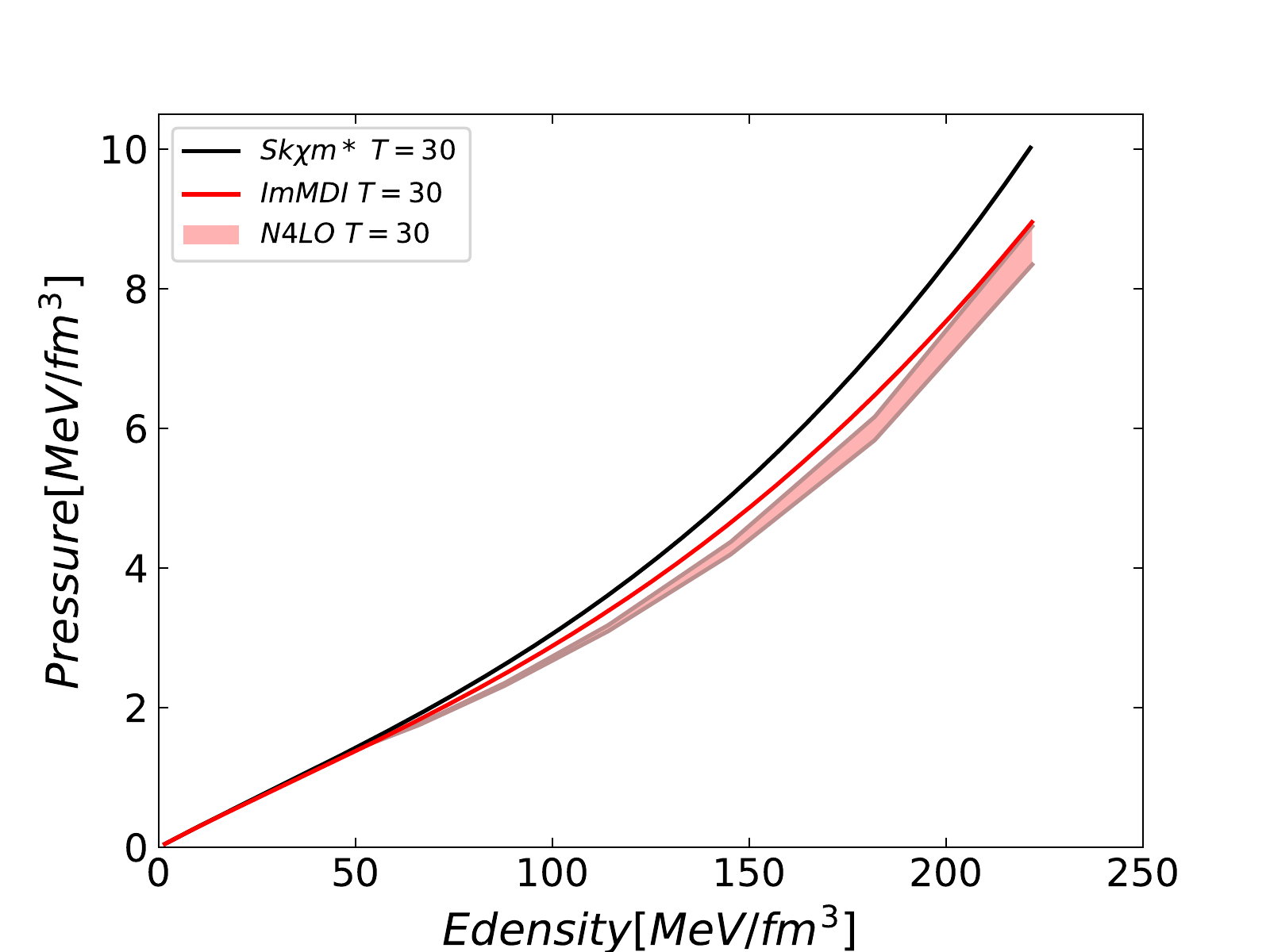}
}
\caption{\label{fig:chptvsXu}
Comparison of the N4LO ChPT computation~\cite{Sammarruca:2020urk} and the Skyrme-based 
computation of~\cite{Xu:2019ouo}. While the comparison is reasonable,
if a bound on a beyond-GR ({\it e.g.} post-Newtonian) coefficient is obtained, what systematic error should be adscribed to the model computation? This application suggests having known uncertainty bands even if the EoS is less accurate for certain properties within General Relativity.
Also, having a well-defined band can help model-makers understand the agreement of computed EoS  with certain basic principles.}
\end{figure}

The resulting
EoS~\footnote{We thank prof. Sammarruca for their computer data.}, that we adopt as our low-density bands, based on~\cite{Sammarruca:2020urk}  are shown in figure~\ref{fig:chpt} .

There are many nuclear model-dependent computations, such as~\cite{Xu:2019ouo} based, for example, on Skyrme-type interactions, that corroborate the increase in $E/A$ with temperature. Most computations are reasonably valid in the low-momentum regime, see figure~\ref{fig:chptvsXu}, but the chiral perturbation theory bands that we adopt showcase the necessity of assigning systematic errors to these other computations.
In the future, one might even be able to dispose of the need for a nuclear potential altogether and directly compute the equation of state from low-energy nucleon-nucleon scattering data, further reducing systematics~\cite{Alarcon:2021kpx}.

\newpage
\section{High density pQCD input} \label{pQCD}
\begin{figure}
\resizebox{0.5\textwidth}{!}{
\includegraphics{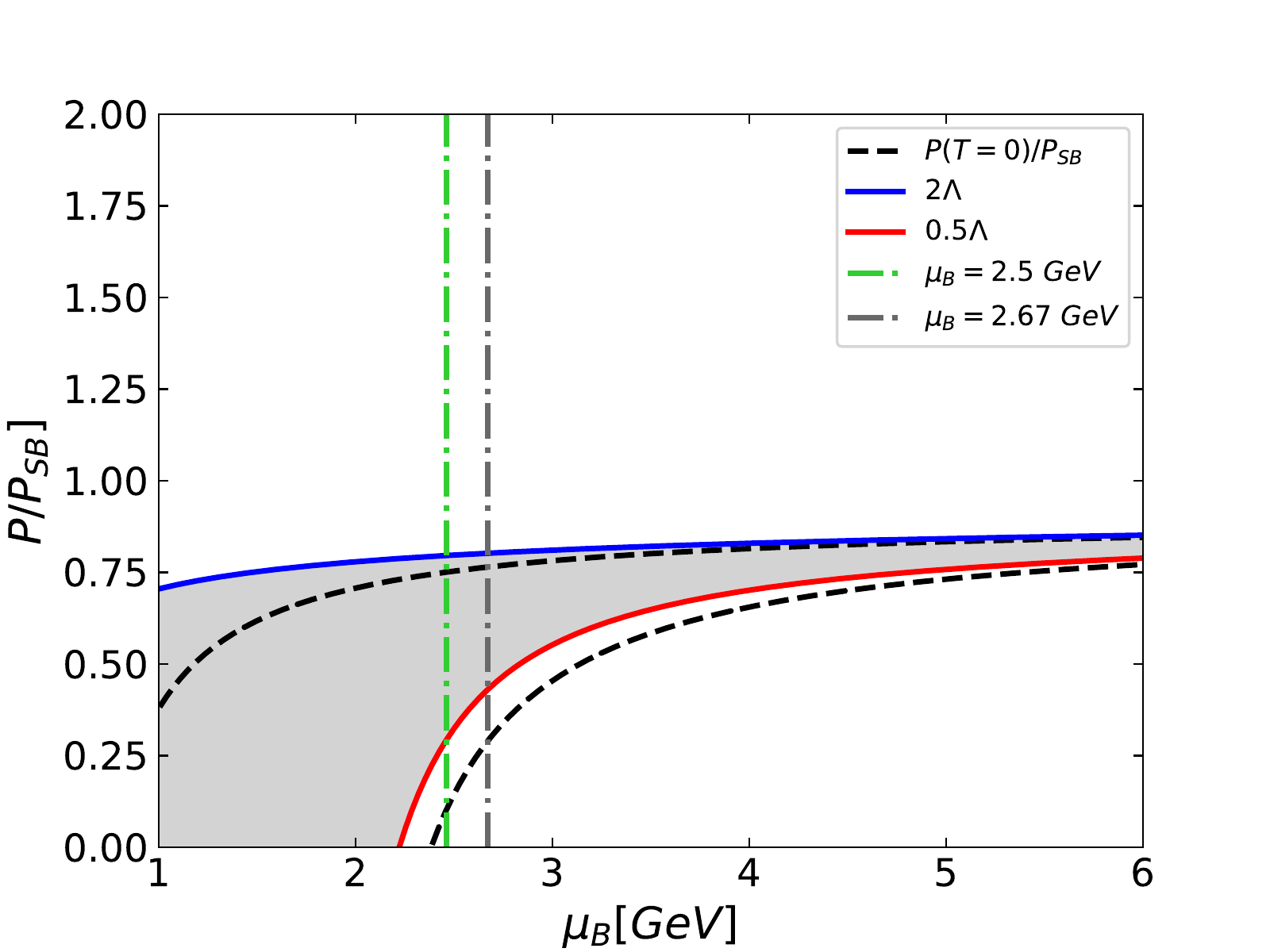}
}
\resizebox{0.5\textwidth}{!}{
\includegraphics{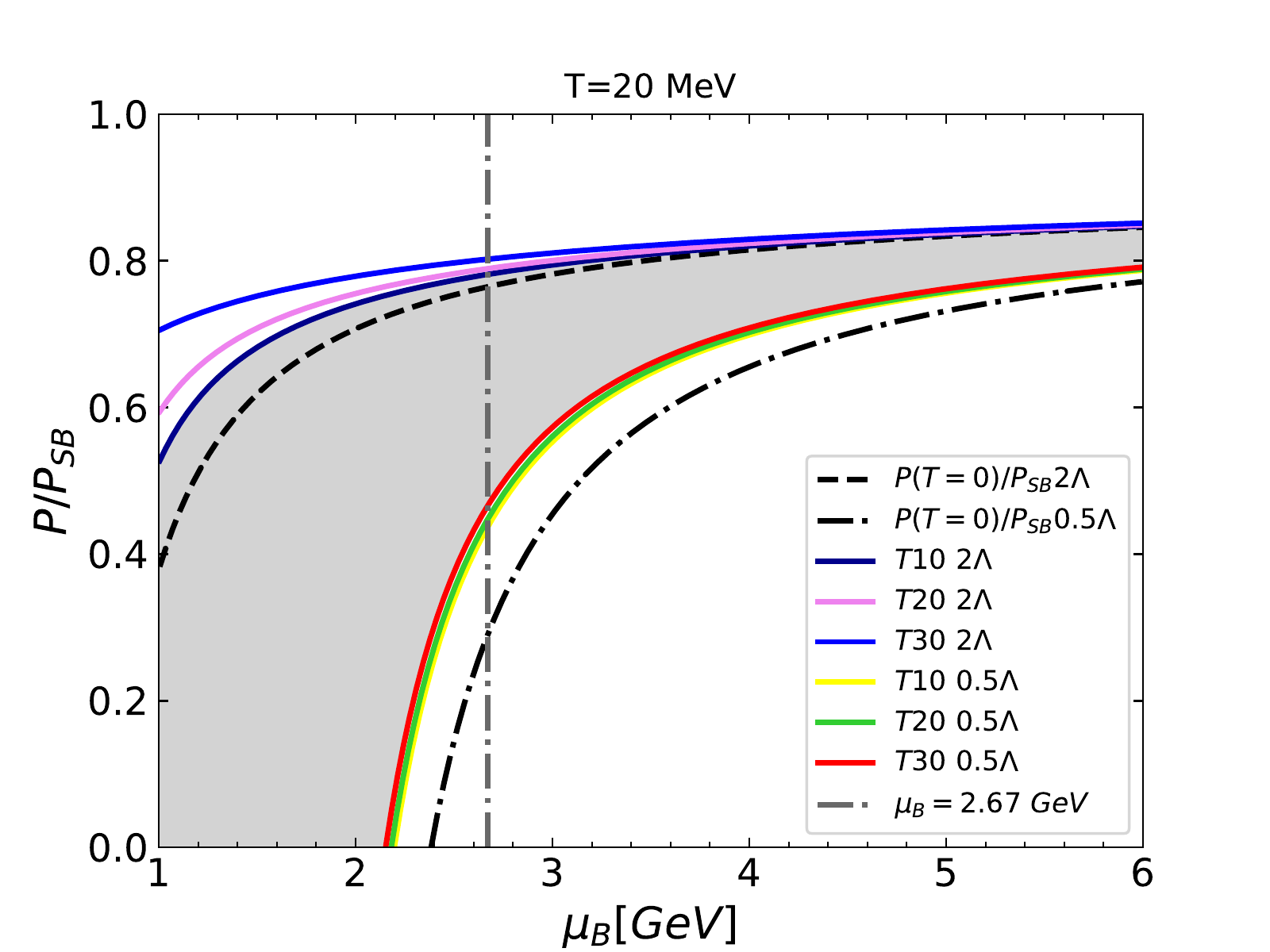}
}
\caption{\label{fig:pQCD30}
Perturbative QCD computation of the pressure of dense matter at the baryon chemical potential indicated in the $OX$ axis and for the temperatures in the legend. In our sets we change the variable $\mu_B$ to an energy density $\varepsilon$ and directly read the equation of state $P(\varepsilon,T)$. The top plot shows the comparison of the $T=30$ MeV EoS band (shaded) with the $T=0$ one (bracketed by the dashed lines). The bottom plot, additionally, displays $T=10$ and 20 MeV lines.
The pressure in both is expressed in units of that of a free-quark gas, given by $P_{SB}=\frac{3\mu^4}{4\pi^2}$  ($\mu= \mu_B/3$ is the quark chemical potential).}
\end{figure}

While hadron interactions are strong and complicated at intermediate momentum, where the coupling is large, for asymptotically large density the smallness of $\alpha_s(\mu)$
allows the deployment of perturbative Quantum Chromodynamics (pQCD).
Thus, all of our EoS terminate inside the high-density band obtained from pQCD computations and shown in Figure~\ref{fig:pQCD30}. 

That band is obtained from perturbative computations at finite density following the Helsinki group work~\cite{Chesler:2019osn,Fraga:2015xha,Vuorinen:2003fs,Kurkela:2016was}
and also~\cite{Kurkela:2009gj} for the $T\to 0$ limit.

Though we have checked our results with the simple approximate parameterization proposed in~\cite{Fraga:2015xha}, our nEoS-T band follows the detailed formulae 
earlier reported in~\cite{Vuorinen:2003fs,Kurkela:2016was} that we have reprogrammed.

The span of the pQCD uncertainty band is an attempt at controlling the systematics of the computation by varying the renormalization scale $\Lambda\in(0.5,2)\bar{\Lambda}$ around the value $\bar{\Lambda}=2\pi\sqrt{T+\frac{(\mu_B/3)^2}{\pi^2}}$. 
The scale variation should be reabsorbed in the nonconformal terms of higher, non-considered orders of perturbation theory without affecting the energy densit or pressure.. However,  a dependence arises which is an artifact of truncating the weak-coupling expansion.
There is some arbitrariness in the choice of scale-uncertainty interval that becomes smaller with a higher order of perturbation theory (the exact series should be independent of the renormalization scale). What this band really does is to orient us in the validity of the perturbative computation: note in the figure how, for chemical potentials below 2 GeV, the uncertainty swells and the computation becomes unusable.

In practice, we vertically slice the band at a baryon chemical potential $\mu_B=2.67$ GeV (this is the nearest point of our grid to the characteristic quark chemical potential that the Helsinki group has been employing, about $\mu_q=0.87$ GeV; for comparison, another line at 2.5 GeV is given). The intersection of that vertical line with the two solid lines (blue and red online), that provide the scale sensitivity in the interval $(0.5\Lambda,2\Lambda)$ act as ``goalposts'' through which any EoS coming from lower density has to pass; those that do not, are automatically discarded. It marks the border between the high-density region (computed in pQCD) and the intermediate density region (interpolated).
This particular point of our grid is chosen as a compromise between the necessity of the uncertainty being small enough but the density being as low as possible to reduce its difference to that of actual neutron stars.

These pQCD EoS entail a three-loop computation of the energy per quark. 
Recently, attempts have been made at narrowing the uncertainty band by means of resummation,
employing the renormalization group~\cite{Kneur:2019tao} but we have been conservative and not incorporated resummed results yet
to avoid introducing some bias or model dependence, at least until further research and independent confirmation reassures as of the procedure as safe.
The same can be said of
Dyson-Schwinger~\cite{Bai:2021non} related approaches, that may be insightful but not yet optimized to provide error bands for beyond-GR searches.

\newpage

It is clear to us that physical neutron stars in General Relativity will not reach such high baryon chemical potentials as described in this section. But there are two reasons to anchor the computation of the EoS in the pQCD one. The first is that modified-gravity theories might actually reach such densities, so we need to provide them with input there. The second, broader and also applicable to GR, is that the EoS of hadron matter at attainable densities is constrained by its necessary behavior at higher ones, due to causality and stability ($\frac{dP}{d\varepsilon}\in[0,1)$).

\newpage

\section{The nEoS sets at finite Temperature}

We are now ready to mount the interpolation. We take the low-density ChPT band and start sampling it from lower to higher values of $n$; after its maximum density is reached,
we continue sampling a broadening band sandwiched between the extreme lines allowed by the conditions of causality $dP/d\varepsilon\leq 0$ and thermodynamic stability $dP/d\varepsilon \geq 0$ between the ChPT and the pQCD bands.

This is accomplished by Von Neumann's rejection, with the program described in \cite{Oter:2019kig}. If at any step in the numeric advance from $(\varepsilon_i,P_i)\to (\varepsilon_{i+1},P_{i+1})$ any of those two conditions is even locally violated, the EoS is rejected and we start anew. 

Thousands of such equations, exploring all the systematics of the ChPT cutoff, the pQCD renormalization scale, and the different computations of the low-density limit, are provided for zero temperature in our website {\tt http://teorica.fis.ucm.es/nEoS}.

In this work, the extension to (small) finite temperature, we adopt a different strategy. Instead of providing multiple samples of the band (all workable EoS compatible with hadron constraints), we make available the extremes of the nEoS bands formed from 
(a) the $N^3LO$ chiral perturbation theory computation at 0, 10, 20, 30 MeV in section~\ref{ChPT};
(b) the pQCD computation described in section~\ref{pQCD}; and (c), the interpolation between both limits  for the corresponding temperatures, satisfying stability and causality.

\begin{figure}[h]
\resizebox{0.5\textwidth}{!}{
\includegraphics{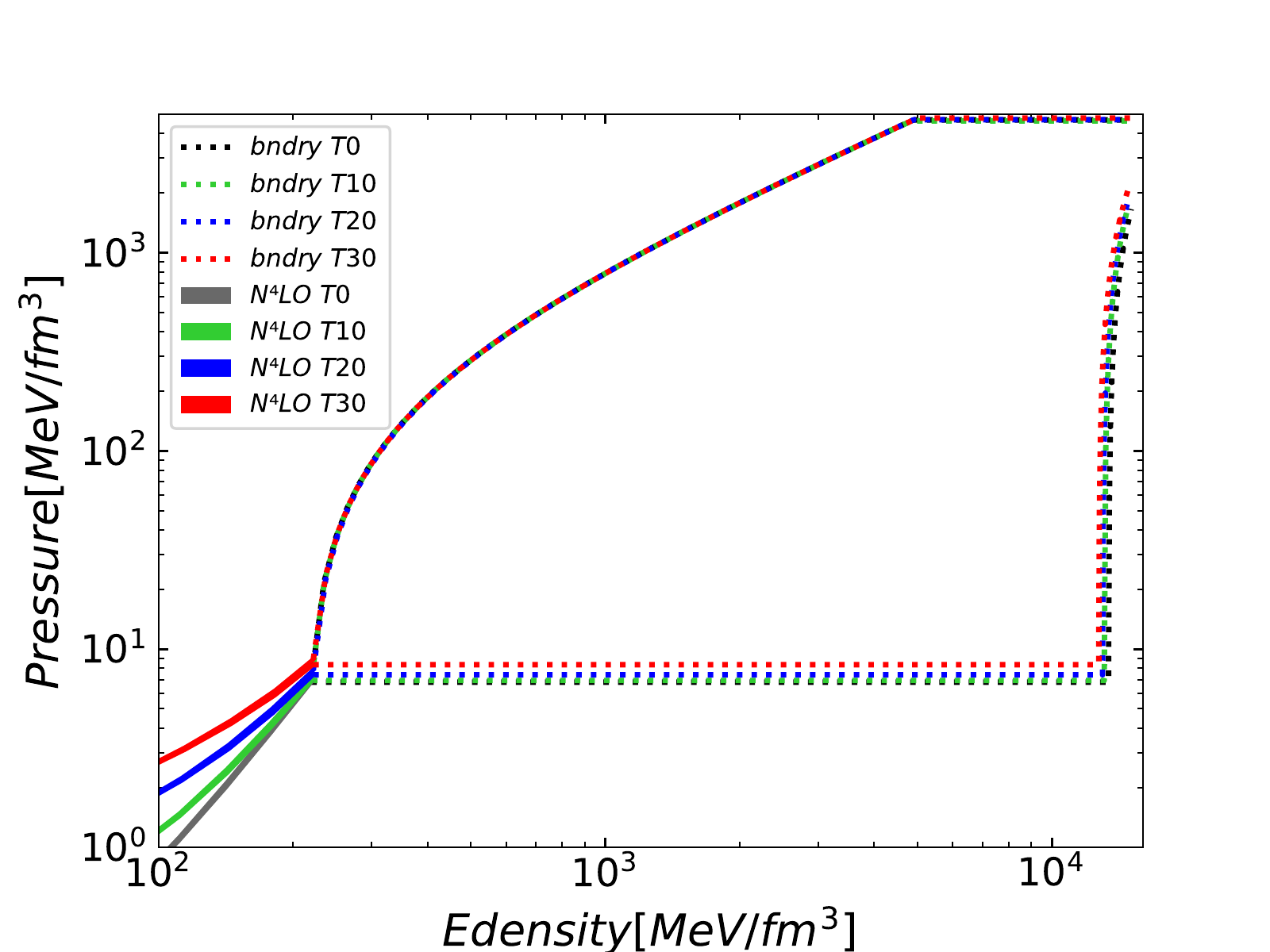}}
\caption{\label{fig:bandT} Comparison of the nEoS-T bands for different temperatures from 0 to 30 MeV. While the low-density equation of state is quite sensitive to even such small temperatures, the intermediate-density bands do not vary much (barely visible in this logarithmic scale)}
\end{figure}

Figure~\ref{fig:bandT} shows the band up to the end of the intermediate interpolated zone (the higher density pQCD part is not show to avoid reducing the scale too much; it is the same as in figure~\ref{fig:pQCD30}).

In addition to our computer server just mentioned,
we will be making these band extremes available through the COMPOSE website
{\tt https://compose.obspm.fr/ } where many other EoS sets for neutron stars can be found. We also provide the geometric mean of the band extremes at largest and lowest pressures, as an example of a typical EoS through the middle of the allowed band.

\begin{figure}
\resizebox{0.5\textwidth}{!}{
\includegraphics{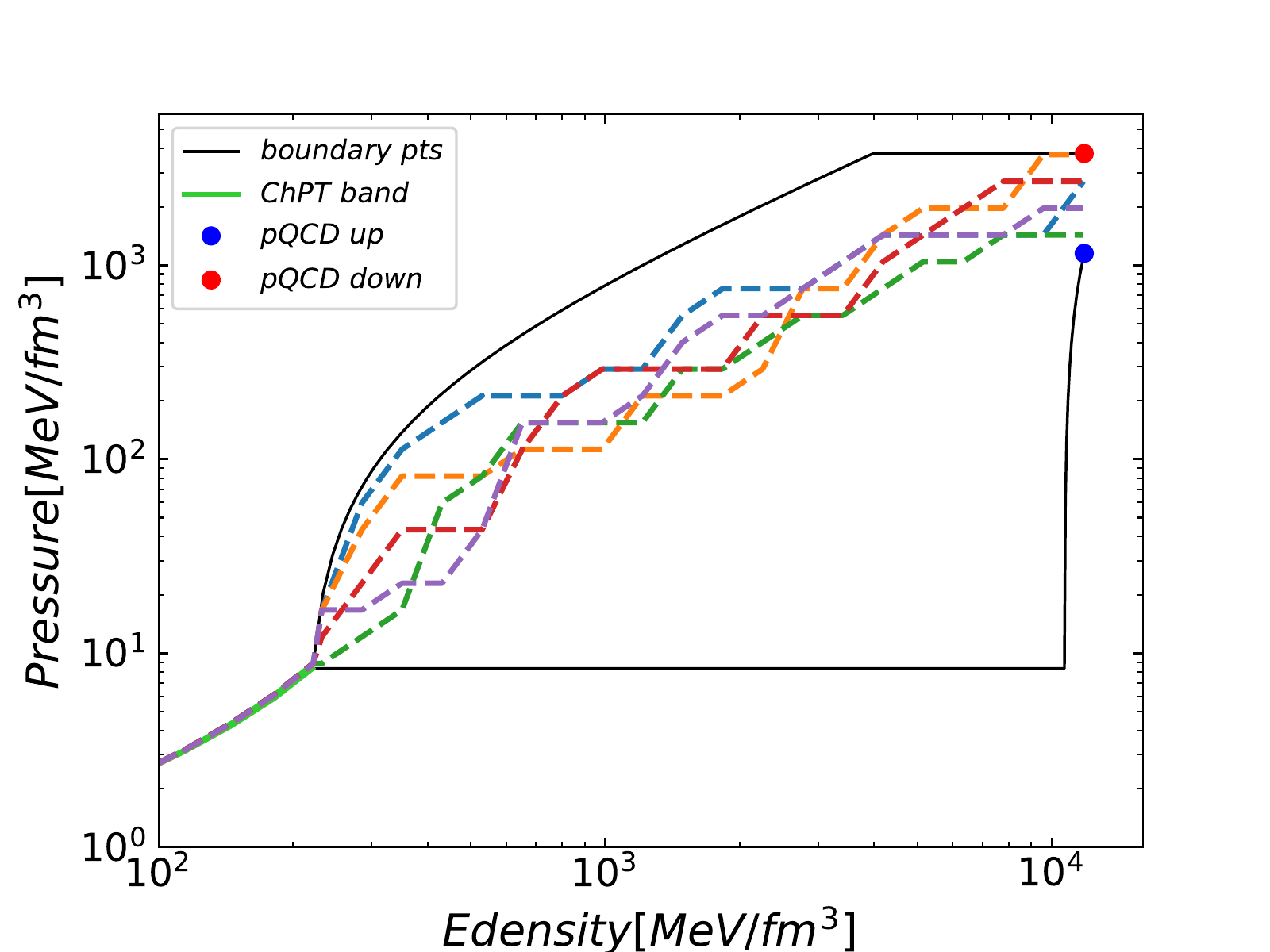}}
\resizebox{0.5\textwidth}{!}{
\includegraphics{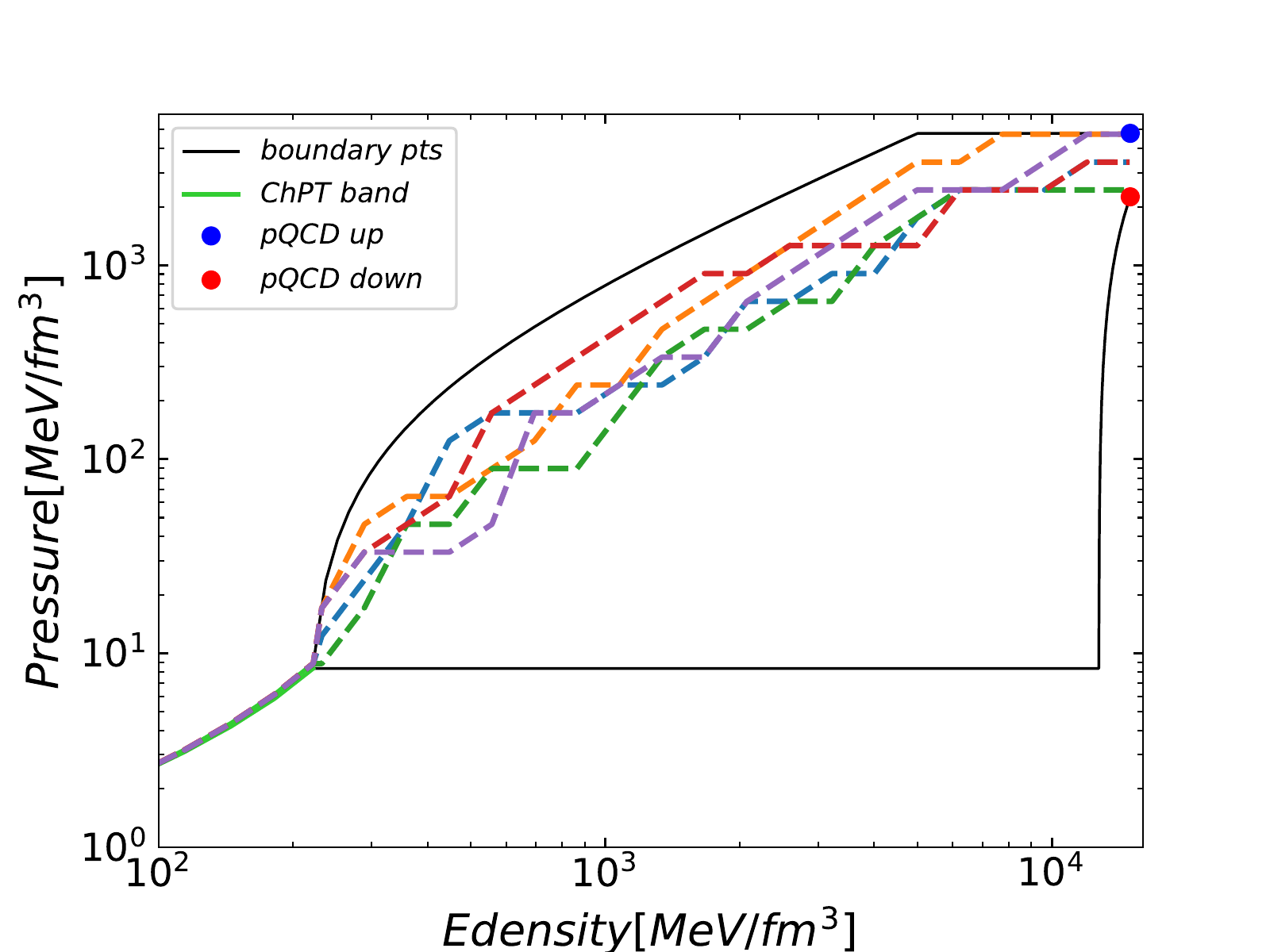}
}
\caption{\label{fig:EosT30} The nEoS-T  band of allowed equations of state in the intermediate density region at finite temperature $T=30$ MeV. The top plot corresponds to $\mu_B=2.5$ GeV, the lower plot to 2.67 GeV. A few allowed equations of state within the band are also shown. }
\end{figure}

Additionally, in this article we show a few examples of allowed EoS that pass through the band in figure~\ref{fig:EosT30}. These examples illustrate a feature of linearly interpolated and stepwise generated EoS: the possibility of having flat sections of zero derivative that can represent first order phase transitions. The longer these flat segments are in the graph, the larger the latent heat of the transition, with the nEoS band introducing an upper bound to the latent heat that is possible in hadron physics~\cite{Lope-Oter:2021mjp}.

\newpage

\section{Comparison to the additive approach}
Often used is the additive (or hybrid) approach~\cite{Janka} that approximates pressure and energy density at finite temperature by the $T=0$ counterpart for cold nuclear matter and an additive thermal correction

\begin{eqnarray} \label{additive}
\epsilon = \epsilon_{\rm cold}+ \epsilon(T)\nonumber \\ \nonumber 
P= P_{\rm cold} + P(T) \\ 
P(T) = (\Gamma-1) \epsilon(T)
\end{eqnarray}
in terms of a simple adiabatic constant $\Gamma$.

This has been recently employed in~\cite{Raithel:2021hye}, together with a more sophisticated so-called $M^*$ parametrization of the relation between $P(T)$ and $\varepsilon(T)$. The authors find that the thermal pressure at $T=20$ MeV is of the same size of the cold-matter pressure at saturation density, but the ratio falls rapidly so that by $n=10n_{\rm sat}$, the thermal pressure is a 10\% correction at most.

In any case, Eq.~(\ref{additive}) entails that the following ratio is constant in the approach,
\begin{equation} \label{checkadd}
\frac{P-P_{\rm cold}}{\varepsilon-\varepsilon_{\rm cold}} = \Gamma-1 =  {\rm constant} 
\end{equation}
an assumption that we here quickly examine. 

\begin{figure}[h]
\resizebox{0.5\textwidth}{!}{
\includegraphics{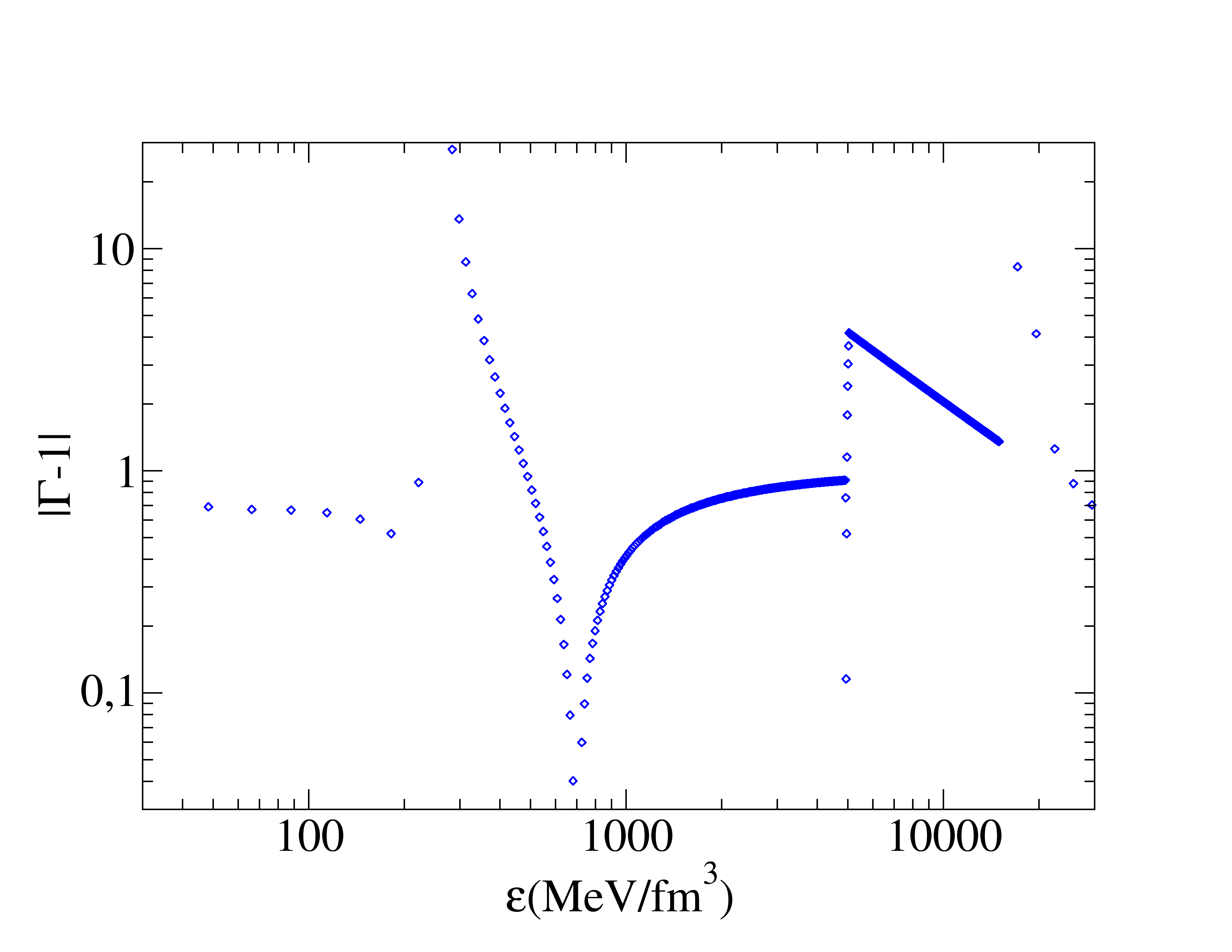}}
\caption{\label{fig:checkadd} Test of the additive approach to thermal effects on the neutron star EoS, plotting the quantity of Eq.~(\ref{checkadd}) for $T=30$ MeV.}
\end{figure}

We have plotted Eq.~(\ref{checkadd}) evaluated for $T=30$ MeV
in figure~\ref{fig:checkadd} following the top of the band (that is the most relevant for maximum mass neutron stars). The plot is quite bumpy, as one can identify the matching points between density regimes around 270 and 15000 MeV/fm$^3$ as well  as the transition between causality-dominated (maximum slope of 1) and stability-dominated (minimum slope of 0) around
5000 MeV/fm$^3$ as the pQCD phase is approached.
Once these nonanalytic features have been discounted (that can presumably also appear
for the one true equation of state if it presents phase transitions between different states of nuclear matter), we can focuse on the ``safe'' regions.

They have very substantial values of the derivative of $|\Gamma-1|$. In the low-density 
regime, the variation of the function plotted, and thus the uncertainty of the additive approach, is of order 20\%, which is reasonable for many applications. 

The change is however of order 100\% along the top line of figure
\ref{fig:bandT} for energy densities in the few hundreds and thousands of MeV/fm$^3$. This is the density region where the core of neutron stars should be in General Relativity and slightly modified theories thereof. 

For the flat stretch of pressure (akin to a first order phase transition) around 10000 MeV/fm$^3$, the ratio $|\Gamma-1|$ falls by about a factor 3, so assuming that it is constant is a poor approximation. This is a density regime, however, that will only be relevant for actual neutron star physics in theories that significantly separate from General Relativity (on the weaker side, as reaching such large densities requires to slow-down collapse). 

So we conclude that the additive approach is reasonable for practical applications at relatively low-densities (still, large compared to $T$) within GR and becomes more dubious in other circumstances.

\section{Discussion}

We have presented an extension of the nEoS sets~\cite{Oter:2019kig} to (small) finite temperature. 
For this we use as input the finite-$T$ chiral computation of~\cite{Sammarruca:2020urk} that determines the low-density EoS of pure neutron matter. At the high-density limit, in contrast, we have used the pQCD computation of~\cite{Kurkela:2016was}
with $\mu_u=\mu_s$, that is, employing $SU(3)$ flavor symmetry; that is, we interpolate between two apparently different regimes as the weak interactions are concerned. 

This is a theoretical choice based on the information at hand at the low-density limit, where pure neutron matter is a good approximation, and trying to stay close to $\beta$-equilibrium at high densities, where the strange quark mass is a correction.  In any case, through the central zone of intermediate densities, the range of pressures allowed at a given density due to the stability-causality band is very large, and this makes the choice of flavor-breaking chemical potential a correction. 
This can be seen, for example, from the expression valid for cold quark matter
\begin{equation}
P(\mu) = a_4 \mu^4 - a_2 \mu^2 m_s^2 +b_2 \mu^2 \Delta^2 +B\dots
\end{equation}
that shows the large $\mu$ expansion, with subleading terms carrying the strange quark mass, the CFL gap $\Delta$ if such asymptotic phase is formed, the ground-state bag constant $B$ or equivalent, etc. It is the first term that dominates the expansion, and it is flavor-blind.
Future work might address beta equilibrium through the entire range of densities. 

\begin{figure}[h]
\resizebox{0.5\textwidth}{!}{
\includegraphics{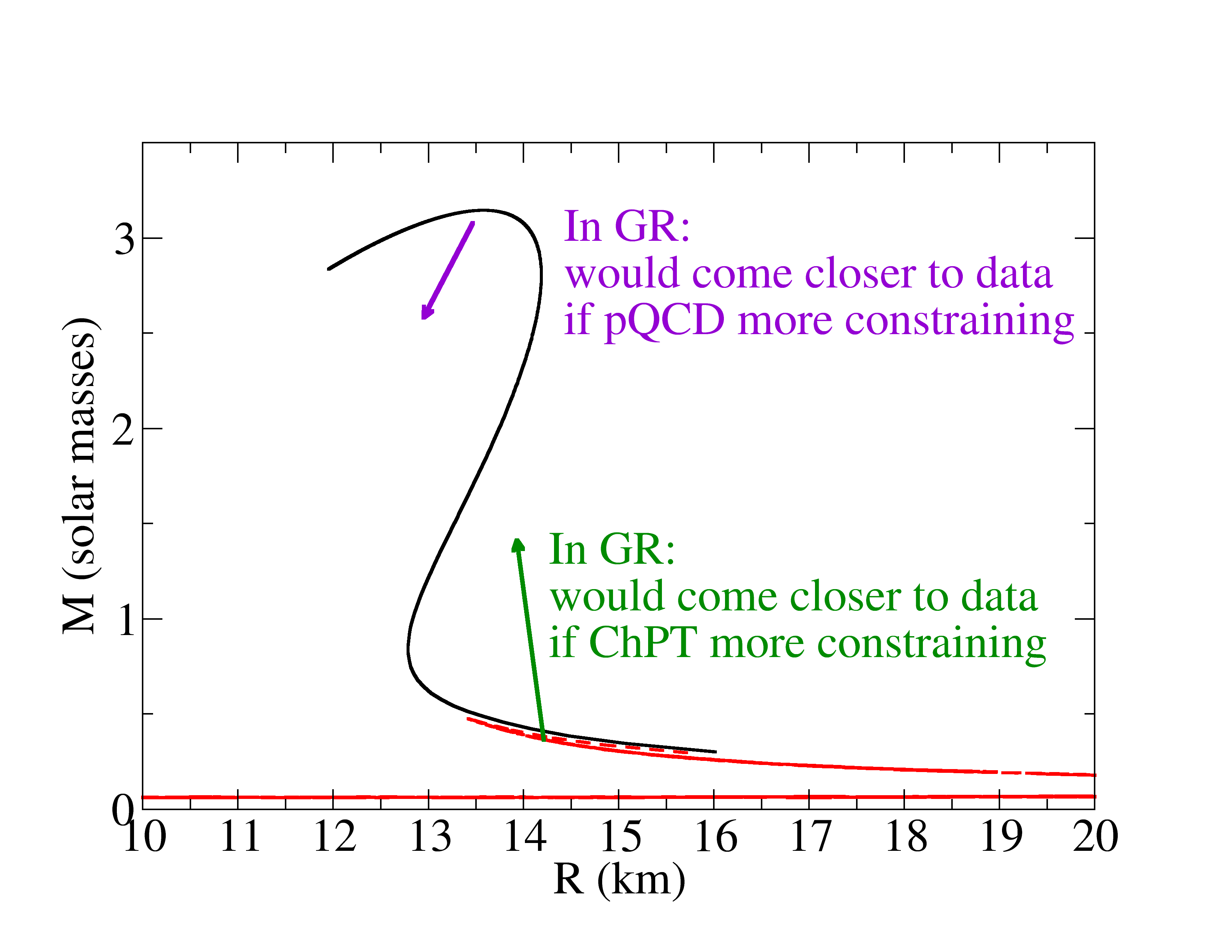}}
\caption{\label{fig:MofR} Mass/radius diagram from numerically solving the Tolman-Oppenheimer-Volkov equations, for the nEoS extreme lines at $T=0$. Basically, both extremes seem excluded in the framework of General Relativity,  meaning that hadron physics computations are less competitive than astrophysical ones for now. However, modifications of GR could bring one or both of these curves back to agree with data, so they should be included in studying modified gravity.}
\end{figure}

\newpage
Perhaps useful to orient the reader as to what hadron physics currently allows at zero temperature is the traditional $M(R)$ plot that we sketch in figure~\ref{fig:MofR} for the usual General Relativity case.
 
 The top, black line corresponds to the stiffest possible EoS allowed by hadronic and theoretical constraints at $T=0$, the lower one (red online)  to the softest. 
 
One could argue that this plot alone would exclude, within General Relativity, the two extreme lines of the nEoS band. The stiffest line reaches masses above 3 $M_\odot$, in contrast to  even optimistic approaches that do not foresee neutron star masses above 2.6 $M_\odot$ from statistical analysis of stellar populations~\cite{Rocha:2021zos}.
The softest line on the other hand does not even reach one solar mass (!). Thus, hadron physics has much room to improve by itself. Of course, if GR is taken for granted, a large swath of the nEoS band must disappear to make the stiffest EoS softer and the softest one, stiffer, as marked in the figure. 

Stronger constraints from the high-density pQCD  side (for example, through adequate resummation of perturbation theory) would affect the former, while extending the ChPT band to higher densities would much affect the second. These improvements do not seem far-fetched and we expect them in the next few years.

In turn, the finite-$T$ EoS is of importance to extract information from the neutron-star like object formed in mergers of two ordinary neutron-stars such as GW170817, that can be hot (due to friction) neutron-matter ephemeral systems supported by fast rotation. They can be used, for example~\cite{Khadkikar:2021yrj}, to bracket the maximum neutron star mass with gravitational wave data.

We have examined temperatures $T=0$ through $T=30$ MeV, following the existing~\cite{Sammarruca:2020urk} ChPT computations. The temperature dependence here is appreciable (figure~\ref{fig:chpt}):  every 10 MeV, the central value of the $N^3LO$ chiral computation is outside the uncertainty band of the neighbooring temperatures. Therefore, $T$ does have to be taken into account into computations involving densities at the nuclear scale.  
The high-density computations, on the other hand, are not too sensitive to $T$, as corresponds to the scale hierarchy $T\ll \mu_B$ (30 MeV {\it versus} 2.6 GeV and above, figure~\ref{fig:pQCD30}). The $T$-dependence becomes remarkable at lower chemical potentials where the perturbative expansion is less reliable anyway and should rather not be used.
Finally, in the intermediate region (figure~\ref{fig:EosT30}), the effect of the temperature is also small when compared with the large swath of parameter space allowed between the low-pressure limit of stability and the high-pressure limit of causality. 

Should the user believe that that small temperature dependence can make a difference in her calculation, our work provides it through the entire density range.

\section*{Acknowledgment}
The authors thank multiple discussions with attendees of the PHAROS workshop: Online repository for the equation of state and transport properties of neutron stars, February 2021.

This project has received funding from the European Union's Horizon 2020 research and innovation programme under grant agreement No 824093; spanish MICINN grants PID2019-108655GB-I00, -106080GB-C21 (Spain);
COST action CA16214 (Multimessenger Physics and Astrophysics of Neutron Stars) and the U. Complutense de
Madrid research group 910309 and IPARCOS.

\newpage


\end{document}